\documentstyle[epsfig,12pt]{article}

\newskip\humongous \humongous=0pt plus 1000pt minus 1000pt

\newif\ifdtup

\newcommand{\GeV}{{\rm\,GeV}}
\newcommand{\MeV}{{\rm\,MeV}}

\newcommand{\ecm}{e\,{\rm cm}}

\def\pl#1#2#3{{\it Phys. Lett. }{\bf B#1~}#3~(19#2)}
\def\np#1#2#3{{\it Nucl. Phys. }{\bf B#1~}#3~(19#2)}

\makeatletter
\newcounter{alphaequation}[equation]
 \def\thealphaequation{\theequation\hbox to
0.6em{\hfil\alph{alphaequation}\hfil}}
\def\eqnsystem#1{
\def\@eqnnum{{\rm (\thealphaequation)}}
\def\@@eqncr{\let\@tempa\relax
\ifcase\@eqcnt \def\@tempa{& & &}
\or \def\@tempa{& &}\or \def\@tempa{&}\fi\@tempa
\if@eqnsw\@eqnnum\refstepcounter{alphaequation}\fi
\global\@eqnswtrue\global\@eqcnt=0\cr}
\refstepcounter{equation}
\let\@currentlabel\theequation
\def\@tempb{#1}
\ifx\@tempb\empty\else\label{#1}\fi
\refstepcounter{alphaequation}
\let\@currentlabel\thealphaequation
\global\@eqnswtrue\global\@eqcnt=0
\tabskip\@centering\let\\=\@eqncr
$$\halign to \displaywidth\bgroup
  \@eqnsel\hskip\@centering
  $\displaystyle\tabskip\z@{##}$&\global\@eqcnt\@ne
  \hskip2\arraycolsep\hfil${##}$\hfil&
  \global\@eqcnt\tw@\hskip2\arraycolsep
  $\displaystyle\tabskip\z@{##}$\hfil
  \tabskip\@centering&\llap{##}\tabskip\z@\cr}

\def\endeqnsystem{\@@eqncr\egroup$$\global\@ignoretrue}
\makeatother

\makeatletter
\renewcommand{\section}{\@startsection{section}{1}{0em}{-\baselineskip}%
{0em}{\large\bf}}
\makeatother

\newcommand{\mycaption}[1]{\caption{{\em #1}}} 
\makeatletter
\renewcommand{\fnum@table}{\bf\tablename~\thetable}
\renewcommand{\fnum@figure}{\bf\figurename~\thefigure}
\makeatother

\begin{document}
\begin{titlepage}
\begin{center}
\today  \hfill IFUP--TH 4/97 \\
~{} \hfill LBNL--39946 \\
~{} \hfill UCB--PTH--97/06  \\

\vskip .25in

{\large \bf Consequences of a U(2) Flavour Symmetry}\footnote{This
work was supported in part by the Director, Office of
Energy Research, Office of High Energy and Nuclear Physics, Division of
High Energy Physics of the U.S. Department of Energy under Contract
DE-AC03-76SF00098, in part by the National Science Foundation under
grant PHY-95-14797 and in part by the U.S. Department of Energy under
contract DOE/ER/01545--700.
This work was partially supported by the ``Beyond the Standard Model"
TMR Network under the EEC contract No. ERBFMRX-CT960090.}

%

\vskip 0.3in

Riccardo Barbieri$^1$,
Lawrence J. Hall$^2$ and Andrea Romanino$^1$

\vskip 0.1in

{{}$^1$ \em Physics Department, University of Pisa\\
     and INFN, Sez.\ di Pisa, I-56126 Pisa, Italy}

\vskip 0.1in

{{}$^2$ \em Department of Physics and
     Lawrence Berkeley National Laboratory\\
     University of California, Berkeley, California 94720, USA}

\end{center}

\vskip .3in

\begin{abstract} \noindent While solving the supersymmetric flavour
problem, a U(2) flavour symmetry might be at the origin of the pattern
of fermion masses and mixings. The consequences of this hypothesis are
spelled out concerning the parameters of the CKM matrix and their
observability in $B$- and $K$-physics. The relevance of searching for
lepton flavour violation and for the electric dipole moments of the
electron and the neutron is also emphasized.  \end{abstract}

\end{titlepage}

\section{} \label{sec:intro}

Flavour physics is a promising area for discoveries and surprises. On
the experimental side, a number of significant observables are likely
to be measured in the coming years: CP-asymmetries in $B$-decays are an
example, which is obvious but, we hope, far from unique, as indicated
below. On the theoretical front, a qualitative and quantitative
explanation of the pattern of fermion masses and mixing angles is
still elusive, despite considerable effort.

If nature is (approximately) supersymmetric, the flavour problem
acquires a new aspect. In supersymmetric theories there are mass and
interaction matrices for the squarks and sleptons, leading to a richer
flavour structure. In particular, if fermions and scalars of a given
charge have mass matrices which are not diagonalized by the same
rotation, new mixing matrices, $W$, occur at gaugino and higgsino
vertices. In turn, the Flavour Changing Neutral Current (FCNC) and/or
CP-violating phenomena induced by these new interactions must not
violate the corresponding experimental bounds.

In a given supersymmetric theory, these different aspects of the
flavour problem may or may not be related to each other. A prevailing
attitude, so far, has been to consider the flavour structure of the
scalar mass matrices as being unrelated to the one of the fermion mass
matrices: the flavour blind mechanism that generates the supersymmetry
breaking scalar masses results in a sufficient degeneracy of the
sfermions to keep the FCNC and CP-violating phenomena under
control. Without discussing here in which cases this physical
consequence is actually justified, in this work we take, as others
have done~\cite{list}, an opposite viewpoint: the flavour structure of
the mass matrices of the fermions and of the scalars are related to
each other by a symmetry principle. It is this principle which is at
the origin of the pattern of fermion masses and mixings and is, at the
same time, responsible for the sufficient suppression of FCNC and
CP-violating phenomena.

As argued in previous works~\cite{b1,b2}, a U(2) flavour symmetry
in which the lighter two generations transform as a doublet and the
third generation as a trivial singlet, like the Higgs fields, emerges
as a candidate to fulfill this role. There are simple reasons for
that. In the limit of unbroken U(2), only the third generation of
fermions can acquire a mass, whereas the first two generations of
scalars are exactly degenerate. While the first property is not a bad
approximation of the fermion spectrum, the second one is what one
needs to keep under control FCNC and CP-violating
phenomena. Furthermore, the rank 2 of U(2) allows a two
step breaking pattern
\begin{equation}
\mbox{U(2)} \stackrel{\epsilon}{\rightarrow} \mbox{U(1)}
\stackrel{\epsilon'}{\rightarrow} 0, 
\label{u2br}
\end{equation}
controlled by two small parameters $\epsilon$ and $\epsilon'$, to be at
the origin of the generation mass hierarchies $m_3\gg m_2\gg m_1$ in the
fermion spectrum. Although it is natural to view U(2) as a subgroup of
U(3), the maximal flavour group in the case of full intra-family gauge
unification, U(3) will be anyhow strongly broken to U(2) by the large
top Yukawa coupling.

\section{} \label{sec:U(2) break}

To fully exploit the consequences of the U(2) flavour symmetry, its
breaking pattern must be specified more precisely. As mentioned, the
three generations of matter fields $\psi$ transform as $\bf 2\oplus
1$,
\[
\psi=\psi_a\oplus\psi_3.
\]
Taking the Higgs bosons to be flavour singlets, the Yukawa
interactions transform as: $(\psi_3 \psi_3)$, $(\psi_3 \psi_a)$,
$(\psi_a \psi_b)$.  Hence the only relevant U(2) representations for
the fermion mass matrices are $1$, $\phi^a$, $S^{ab}$ and $A^{ab}$,
where $S$ and $A$ are symmetric and antisymmetric tensors, and the
upper indices denote a U(1) charge opposite to that of
$\psi_a$. $\phi^a$, $S^{ab}$ and $A^{ab}$ can be viewed as ``flavon''
fields.

We make the simplifying assumption that each of these fields
participate in only one stage of the symmetry breaking in
(\ref{u2br}). Since $A^{ab}$ alone would break U(2) down to SU(2),
whereas it would break U(2) completely in association with $\phi^a$
and/or $S^{ab}$, it can only participate in the last stage of breaking
in (\ref{u2br}): $\mbox{U(1)}\rightarrow 0$. Therefore,
$A^{12}=-A^{21}={\cal O}(\epsilon')$. On the other hand, to account
for $|V_{cb}|\simeq m_s/m_b$ in term of a unique parameter $\epsilon$,
both $\phi^a$ and $S^{ab}$ must participate in the first stage of
breaking in (\ref{u2br}): 
$\mbox{U(2)}\rightarrow \mbox{U(1)}$\footnote{Without
introducing unnatural coefficients and/or cancellations, the two other
alternatives, $S^{ab} = {\cal O}(\epsilon)$ and $\phi^a = {\cal O}
(\epsilon')$, or viceversa,
would lead to $|V_{cb}| \approx |V_{us}| m_s/m_b$ or $|V_{cb}| \approx
(m_s/m_b)^{1/2}$, respectively.}.
Hence,
in the basis where $\phi^2={\cal O}(\epsilon)$ and $\phi^1=0$,
$S^{22}={\cal O}(\epsilon)$ and all other components of $S$ vanish --
if they were non-zero they would break U(1) at order $\epsilon$, which
is excluded by (\ref{u2br}). We are thus led to Yukawa matrices in
up, down and charged lepton sectors of the form:
\begin{equation}
\left( \begin{array}{ccc}
0 & \epsilon' & 0 \\
-\epsilon' & \epsilon & \epsilon \\
0 & \epsilon & 1
\end{array} \right).
\label{lambda}
\end{equation}
All non vanishing entries have unknown coefficients of order unity,
while still keeping $\lambda_{12}=-\lambda_{21}$. With $\epsilon
\simeq 0.02$ and $\epsilon' \simeq 0.004$, such a pattern agrees
qualitatively well with the observed quark and lepton masses and
mixings, with a few exceptions which can be understood in terms of the
composition of the Higgs which couple to the $D$/$E$ sectors and of the
intra-generation structure of the Yukawa couplings~\cite{b2}.

The mass and interaction matrices for the scalars arising from
supersymmetry breaking can be discussed along similar lines. We assume
that the same representations, $\phi^a$, $S^{ab}$ and $A^{ab}$, which
play a role in the fermion Yukawa sector, are also relevant in the
description of flavour breaking in the scalar sector. In this way, it
is immediate to see that the ``$A$-terms'' have the same structure of
the Yukawa matrices in (\ref{lambda}), whereas the only terms of
numerical relevance in the scalar mass matrices are those ones linear
in $\phi^a$, $(\phi^a)^\dagger$, $\phi^a (\phi^b)^\dagger$ and
$S^{ab}(S^{bc})^\dagger$.  Hence, the resulting mass matrices have the
form
\begin{equation}
m^2 = \left( \begin{array}{ccc}
m_1^2 & 0 & 0 \\
0 & m_1^2 (1 + \epsilon^2) & \epsilon {m_4^2}^* \\
0 &  \epsilon m_4^2 &  m_3^2
\end{array} \right)
\label{m^2}
\end{equation}
where $m_1$, $m_3$ and $m_4$ are masses of the order of the
supersymmetry breaking scale.

\section{} \label{sec:Yukawa}

The CKM matrix has many possible forms, as there are many ways to choose
the three Euler angles. 
For example the original choice of Kobayashi and Maskawa took the form
$V = R_{23}(\theta_{23})  R_{12}(\theta_{12}) P(\delta)  
R^T_{23}(\theta'_{23})$, where $R_{ij}(\theta_{ij})$ is a $2 \times 2$ rotation
in the $ij$ plane by angle $\theta_{ij}$. $P$ is a diagonal phase matrix with
non-zero entries $(1, 1, e^{i \delta})$.
To appreciate the relationship between the CKM
matrix and the quark masses there is a preferred choice for the Euler
angles: the larger terms (perturbations) in the quark mass matrices
should be diagonalized first. This suggests consideration of the form
$V = R_{12}(\theta^U_{12})  R_{23}(\theta_{23}) P(\phi)  
R^T_{12}(\theta^D_{12})$,
where the diagonal phase matrix has entries $(e^{-i\phi},1,1)$.
For the case that the 13, 31 and 11 entries
vanish, and the 12 and 21 entries have equal magnitudes, as in
(\ref{lambda}), a perturbative diagonalization requires first a 23
rotation, then a 12 rotation giving~\cite{HR}
\begin{equation}
V = \left( \begin{array}{ccc}
c_{12}^D + s_{12}^D s_{12}^U e^{i\phi}   
& s_{12}^D - s_{12}^U e^{i \phi} & -s_{12}^U s \\
s_{12}^U -s_{12}^D e^{i \phi} & c_{12}^D  e^{i \phi} + s_{12}^U s_{12}^D
&  s \\ s_{12}^D s                    &  -c_{12}^Ds            
& e^{-i\phi} 
\end{array} \right)
\label{CKM}
\end{equation}
where
\begin{equation}
s_{12}^D = 
\sqrt{{m_d \over m_s}} \left( 1 - {m_d \over 2 m_s} \right),
\label{12D}
\end{equation}
\begin{equation}
c_{12}^D = \sqrt{1 - (s_{12}^D)^2},
\label{c12}
\end{equation}
and
\begin{equation}
s_{12}^U =  
\sqrt{{m_u \over m_c}},
\label{12U}
\end{equation}
with $m_u$ and $m_c$, as $m_d$ and $m_s$, renormalized at the same
scale. The biggest errors in (\ref{CKM}) are in
$V_{ub}$ and $V_{td}$, of relative order $s(m_d m_c/m_u m_s)^{1/2}$
\footnote{The precise values of these correction depends on the
Yukawa matrix elements $\lambda_{23}^{U,D}$ and
$\lambda_{32}^{U,D}$. Taking
$\left|\lambda_{23}/\lambda_{32}\right|^{U,D} = 1$ at the unification
scale and barring significant cancellations in the determination of
$s$, the largest correction occurs in $\left| {V_{ub}/V_{cb}} \right|=
\sqrt{m_u/m_c}$, and
ranges from about 1 to 7\%, depending on relative phases.}  and
$s$ respectively. 

At first sight this is a cumbersome form for the CKM matrix, since
$V_{us}$ contains two terms with a relative phase: $V_{us} \equiv s_c
e^{-i\beta} = s_{12}^D - s_{12}^U e^{i\phi}$. However, this is nothing
other than the well-known unitarity triangle, with $\phi = \alpha$ and
$\beta$ the usual angles:
\begin{eqnsystem}{albega}
& \displaystyle \alpha  \equiv 
\arg\left(-\frac{V_{tb}^*V_{td}^{}}{V_{ub}^*V_{ud}^{}}\right) 
\approx \phi & \\
& \displaystyle \beta  \equiv 
\arg\left(-\frac{V_{cb}^*V_{cd}^{}}{V_{tb}^*V_{td}^{}}\right) 
\approx \arg\left(1-\frac{s^U_{12}}{s^D_{12}}e^{-i\phi}\right) & \\
& \displaystyle \gamma \equiv
\arg\left(-\frac{V_{ub}^*V_{ud}^{}}{V_{cb}^*V_{cd}^{}}\right) 
\approx \pi-\alpha-\beta. &
\end{eqnsystem}
This can be seen by rephasing (\ref{CKM}) into the form 
\begin{equation}
V \approx \left( \begin{array}{ccc}
1   & s_c & -s_{12}^U s \\
-s_c &   1 &  se^{-i(\alpha+ \beta)} \\ 
s_{12}^D s    &  -s e^{i \beta}   &e^{-i\alpha} 
\end{array} \right)
\label{nCKM}
\end{equation}
The orthogonality of the first and third columns, which gives the usual
unitarity triangle, gives $s(s_ce^{-i\beta} - s_{12}^D +
s_{12}^U e^{i\alpha})=0$. In this form it is clear that $\beta$ is
important for much kaon physics: for example, the $W$-box contribution
to $\epsilon_K$ is proportional to $\sin(2\beta)$.

	The simple pattern of U(2) breaking implies that, given the masses of
the four light quarks, all CKM matrix elements are known in terms of just
two free parameters, $s=|V_{cb}|$ and $\phi$. Using the values for
$|V_{us}|$, $|V_{cb}|$ and $|V_{ub}/V_{cb}|$ listed in Table 1, we can
predict $|V_{td}/V_{ts}|= (m_d/m_s)^{1/2}$ and the CP-violating phase
$\phi$, as well as the angles $\beta$ and $\gamma$ of the  unitarity
triangle.

For $m_c$ and $m_s$, we use the values listed in
Table~\ref{tab:inputs}, whereas the ratios of the light quark
masses, $m_u/m_d$ and $m_d/m_s$ require a discussion.  Second order
chiral perturbation theory for the pseudoscalar meson masses determines, 
to a remarkable accuracy, the combination~\cite{leutwyler}
\begin{equation}
Q = {{m_s \over m_d} \over \sqrt{1- {m_u^2\over m_d^2}}} = 22.7 \pm 0.08.
\label{Q}
\end{equation}
 Additional
assumptions, plausible but not following from pure QCD, lead
to~\cite{leutwyler}
\begin{equation}
m_u/m_d = 0.553 \pm 0.043.
\label{u}
\end{equation}
Finally, a better determination of the scale of the light quark masses
is possible if use is made of the SU(5) relations, valid at the
unification scale,
\begin{equation}
\label{SU5}
m_b = m_\tau,\qquad
m_d m_s = m_e m_\mu
\end{equation}
as illustrated in~\cite{b2}. Given the different level of uncertainty
and/or assumptions in these equations, we describe the results of 4
``combined fits'' with different inputs for the light quark mass
ratios: i) (\ref{Q}) only; ii) (\ref{Q}) and (\ref{u}); iii) (\ref{Q})
and (\ref{SU5}); iv) (\ref{Q}), (\ref{u}) and (\ref{SU5}).

\begin{table}
\centering
\renewcommand{\arraystretch}{1.1}
\vskip 0.25in
\begin{tabular}{||c|c||c|c||}
\hline
$(m_s)_{1\GeV}$ & $(175 \pm 55)\MeV $ & $|V_{us}|$ & $0.221 \pm
0.002$ \cr 
$(m_c)_{m_c}$ & $(1.27 \pm 0.05)\GeV $ & $|V_{cb}|$ & $0.040 \pm
0.003$
\cr 
$(m_b)_{m_b}$ & $(4.25 \pm 0.15)\GeV $ & $|V_{ub}/V_{cb}|$ & $0.08 \pm
0.02$ \cr 
$(m_t)_{m_t}$ & $(165 \pm 10)\GeV $ & $\alpha_s (M_Z)$ & $0.117 \pm
0.006$ \cr
\hline
\end{tabular}
\mycaption{Values of the parameters used in the text.}
\label{tab:inputs}
\end{table}

Treating all errors as ``gaussian'', at 90\% C.L., we obtain the
results shown in figs.~1. In fig.~1a, the boundary obtained from the
unitarity constraints on a general parametrization of the $V_{\rm
CKM}$ is also shown. No similar boundary is given in figs.~1b, 1c,
since unitarity alone does not limit the CKM phase $\phi$. It is
possible, on the contrary to constrain $\phi$, if one includes also
the CP violating parameter in $K$ physics, $\epsilon_K$, and the $B_d$
mixing mass, as in most Standard Model fits. Using for the standard
quantities which parametrize QCD uncertainties, $B_K = 0.8 \pm0.2$ and
$\sqrt{B}f_B = (200 \pm 40) \MeV$, the result in fig.~1d is obtained.
As explained below, however, this last plot does not have general
validity.

\begin{figure}
\centering
\epsfig{file=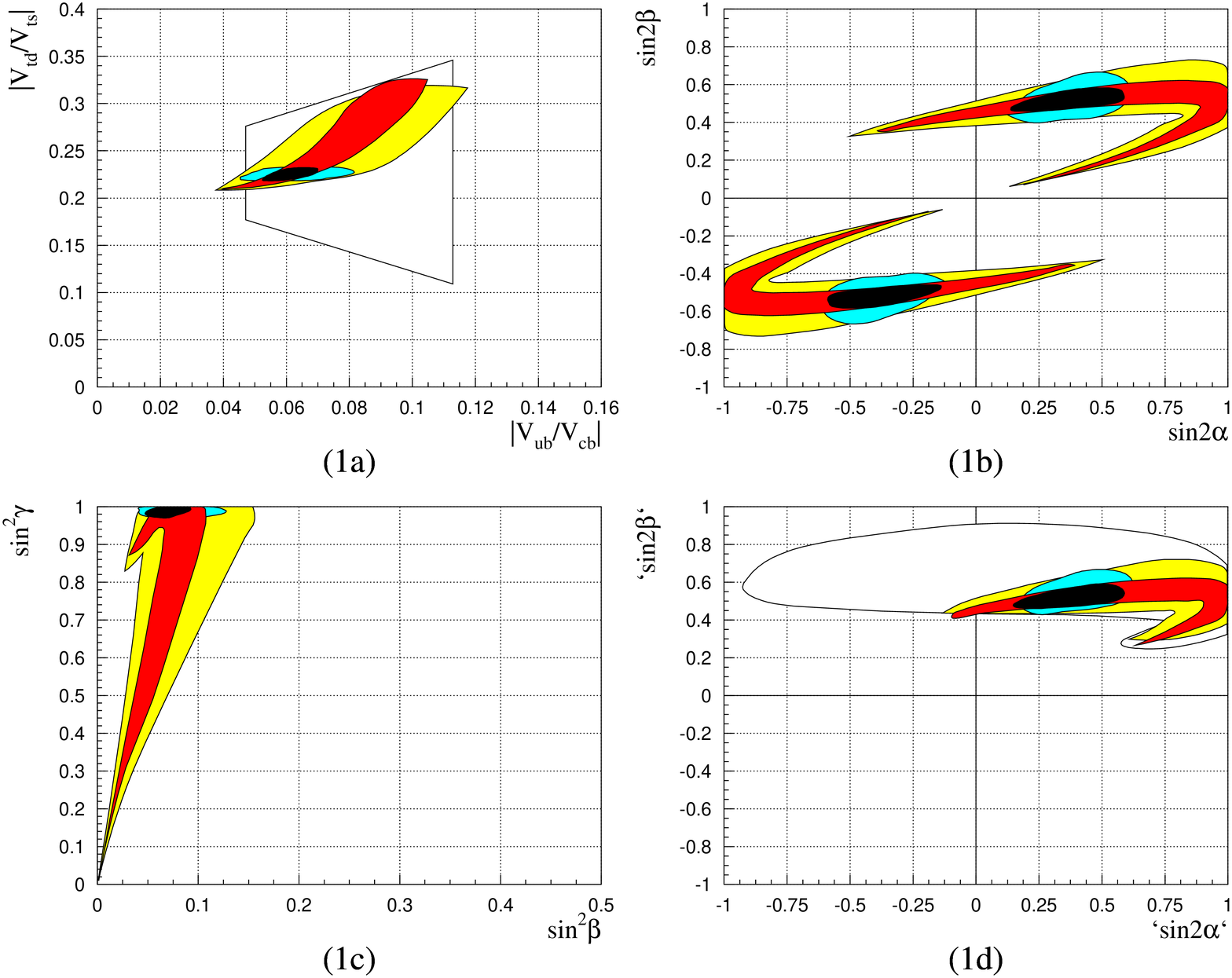,width=\textwidth}
\mycaption{$90\%$ C.L. contours from the four combined fits defined in
the text: i) larger, lighter area; ii) smaller, lighter area; iii)
larger, darker area; iv) smaller, darker area. In figs.\ 1a and 1d
also shown is the contour for a general parametrization of the CKM
matrix. Fig.\ 1d is the only one to include $\epsilon_K$ and $\Delta
m_{B_D}$ among the inputs.}
\label{fig:4fig}
\end{figure}

\section{} \label{sec: W}

The diagonalization of the matrix (\ref{m^2}) leads to highly
degenerate scalars of the first two generations, $(m^2_2 - m^2_1)/m^2
\approx \epsilon^2 \approx 10^{-3}$, with $m^2$ an average scalar mass
squared. On the contrary, the scalars of the third generation are
likely to have masses very different from their first and second
generation partners, $(m^2_3 - m^2_{1,2})\approx m^2$.

By going to a diagonal basis both for (\ref{lambda}) and (\ref{m^2}),
mixing matrices are generated in the gaugino interactions:
$(\bar{f}_{L,R} W_{L,R} \tilde{f}_{L,R}) \tilde{g}$ for the fermions of given
chirality, L or R, and their superpartners. By appropriate phase
redefinitions of the fermion and scalar fields, while keeping the form
(\ref{lambda}) of $V_{\rm CKM}$ and the mass eigenvalues real and
positive, it is possible to write the six matrices $W^{U,D,E}_{L,R}$
for $U$-quarks, $D$-quarks and charged leptons in terms of two new parameters
each, $s^{U,D,E}_{L,R}$ and $\gamma^{U,D,E}_{L,R}$
\begin{equation}
W^{U,D,E}_{L} = \left( \begin{array}{ccc}
c_{12}   & - s_{12} & s_{12} s_L \\
s_{12}  & c_{12}    & - s_L \\
0       & s_L e^{i\gamma_L} & e^{i\gamma_L}
\end{array} \right)^{U,D,E}
\label{WL}
\end{equation}
\begin{equation}
W^{U,D,E}_{R} = \left( \begin{array}{ccc}
c_{12}   &  s_{12} & -s_{12} s_R \\
- s_{12}  & c_{12}    & - s_R \\
0       & s_R e^{i\gamma_R} & e^{i\gamma_R}
\end{array} \right)^{U,D,E}
\label{WR}
\end{equation}
where $s_{12}^{U,D}$ have already been defined and, in analogy with
them, $s_{12}^E = \sqrt{m_e/m_\mu}$. The parameters $s^{U,D,E}_{L,R}$
are all of order $\epsilon$. In this basis, the L-R mixings induced by
the $A$-terms are still non diagonal and complex. They can, however, be
treated as a perturbation except, maybe, for the
$\tilde{t}_L-\tilde{t}_R$ mixing, which, in any event, does not alter
the mixing matrices (\ref{WL}) and (\ref{WR}).

As already mentioned, via loops of supersymmetric particles, the
$W$-matrices give rise to new FCNC and CP-violating phenomena. A close
inspection shows that the most important effects occur in
$\epsilon_K$, in $B-\bar{B}$ mixing, in the Electric Dipole Moments
(EDM) of the electron and the neutron and, finally, in $\mu
\rightarrow e \gamma$ and $\mu \rightarrow e$ conversion in atoms. In
the case of $\epsilon_K$ and $B-\bar{B}$ mixing one obtains effects
comparable to those present in the Standard Model. On the other hand,
the effects in the dipole moments and in the Lepton Flavour Violating
processes are at the level of the present experimental limits. The
calculation of some typical, although partial, contributions to these
observables gives in fact:%
\footnote{In equations (17) and (18) we consider the photino contribution.
For the EDM of the $u$-quark one has $d(u) \approx
8 d(d) (v_1/v_2)^2 (\omega^U/\omega^D)
(\sin(\gamma^U_L-\gamma^U_R)/\sin(\gamma^D_L-\gamma^D_R))$.}
\begin{equation}
\epsilon_K \approx 2 \cdot 10^{-3} \left( {500 \GeV \over
m_{\tilde{q}}} \right)^2 (\omega^D)^2 \sin 2\beta
\label{eps}
\end{equation}
\begin{equation}
\Delta m_{B_d} \approx 0.1\,\mbox{ps}^{-1} \left( {500 \GeV \over
m_{\tilde{q}}} \right)^2 \omega^D
\label{deltam}
\end{equation}
\begin{equation}
\mbox{BR} (\mu \rightarrow e \gamma) \approx 2 \cdot 10^{-11}
\left( {100 \GeV \over m_{\tilde{l}}} \right)^4
(\omega^E)^2
\left( {v_2 \over v_1} \right)^2
\label{muegam}
\end{equation}
\begin{equation}
d(e) \approx  6\cdot10^{-27} \ecm
\left( {100 \GeV \over m_{\tilde{l}}} \right)^2
\omega^E \sin(\gamma_L^E - \gamma_R^E)
\left( {v_2 \over v_1} \right)
\label{dee}
\end{equation}
\begin{equation}
d(d) \approx  1 \cdot10^{-26} \ecm
\left( {500 \GeV \over m_{\tilde{q}}} \right)^2
\omega^D \sin(\gamma_L^D - \gamma_R^D)
\left( {v_2 \over v_1} \right)
\label{ded}
\end{equation}
where $d(e)$ and $d(d)$ denote the EDMs of the electron and the
$d$-quark, respectively, 
$\omega^{U,D,E} = (s^{U,D,E}_L s^{U,D,E}_R)/ V_{cb}^2={\cal O}(1)$ and
$v_2/v_1$ is the ratio of the vacuum expectation values of the Higgs
doublets.

In these equations, the effect of the splitting between the first two
generations of scalars is neglected. This is completely justified for
all the observables except $\epsilon_K$, in which case an additional
effect of the same order of (\ref{eps}) is present, still proportional
to $\sin 2\beta$ as (\ref{eps}) and the SM contribution
themselves. The effects are all due to the splitting $(m^2_3 -
m^2_{1,2})$, taken large enough that the GIM suppression can be safely
neglected.  Therefore, $m_{\tilde{q}}$ in (\ref{eps}), (\ref{deltam})
and (\ref{ded}) denotes the mass of the lightest $Q=1/3$ squark,
whereas $m_{\tilde{l}}$ in (\ref{muegam}) and (\ref{dee}) the mass of
the lightest charged slepton, neglecting in both cases the difference
between L- and R-states. The gluino mass in (\ref{eps}),
(\ref{deltam}), (\ref{ded}) and the
photino mass in (\ref{muegam}), (\ref{dee}) are taken equal to
$m_{\tilde{q}}$ and $m_{\tilde{l}}$ respectively. Eq.~(\ref{deltam})
only includes the contribution of the
$(V-A)(V+A)$ 4-quark operator. The $A$-terms are neglected in
(\ref{muegam}), (\ref{dee}), (\ref{ded}), whereas the
``$\mu$-parameter'' is taken equal to the relevant sfermion mass.

In spite of these limitations and uncertainties, on the basis of
(\ref{muegam})--(\ref{ded}), we look with the greatest possible
interest to any improvement of the experimental sensitivity in the
search for $\mu \rightarrow e \gamma$ (or $\mu \rightarrow e$
conversion)\footnote{ Lepton Flavour Violation could be inhibited by
a suitable extension of the  U(2) symmetry. This is not
compatible, however, with intra-family gauge
unification~\cite{barbieri:94a}.} and for the electron and neutron
EDMs.

\section{} \label{sec: CP}

In view of eq.~(\ref{deltam}), also the mixing and CP-violation in the
$B$-system is likely to deviate from the expectation of the SM, as is
the case for the parameter $\epsilon_K$. In particular, the
experimental determination of the CKM parameters will be affected. For
example, this explains why fig.~1d, which employs SM formulae for
$\epsilon_K$ and $\Delta m_{B_d}$ as inputs, is not generally valid.

Precise formulae for the contribution from the gluino box diagram to
the mixing matrix elements in $B_d-\bar{B}_d$ and $B_s-\bar{B}_s$ are
\begin{equation}
M_{12}(B_d) = (s_{12}^D)^2 \left( {-s_L^D s_R^D e^{i(\gamma_L^D +
\gamma_R^D)} M_{LR} +
(s_L^D)^2 e^{2i \gamma_L^D} M_{LL} +
(s_R^D)^2 e^{2i \gamma_R^D} M_{RR}} \right),
\label{m12d}
\end{equation}
\begin{equation}
M_{12}(B_s) =
s_L^D s_R^D e^{i(\gamma_L^D + \gamma_R^D)} M_{LR} +
(s_L^D)^2 e^{2i \gamma_L^D} M_{LL} +
(s_R^D)^2 e^{2i \gamma_R^D} M_{RR}.
\label{m12s}
\end{equation}
where $M_{LR}$, $M_{LL}$, $M_{RR}$, equal between $B_d$ and $B_s$ 
apart from SU(3)-breaking effects, correspond to the $(V-A)(V+A)$,
$(V-A)(V-A)$, $(V+A)(V+A)$ operators respectively. Using the vacuum
insertion approximation and taking equal masses for right-handed and
left-handed down squarks, $M_{LR} = 4 M_{LL} = 4 M_{RR}$.

Notice the relative minus sign in front of the presumably dominant
LR-contributions to $B_d$ and $B_s$. This should lead to a significant
deviation from the SM result (apart from SU(3)-breaking effects)
\begin{equation}
 {M_{12}(B_d) \over M_{12}(B_s)}  = \left| {V_{td} \over
V_{ts}} \right|^2,
\label{ratio}
\end{equation}
as well as to the appearence of two different extra phases in the
ratio of the mixing coefficients in $B_d$ and $B_s$
\begin{equation}
\label{phases}
\left( {q\over p} \right)_{B_d} \equiv \left( {V_{tb}^*V_{td}\over
V_{tb}V_{td}^*}
\right) e^{-2i\phi_{B_d}}, \qquad
\left( {q\over p} \right)_{B_s} \equiv \left( {V_{tb}^*V_{ts}\over V_{tb}
V_{ts}^*}
\right) e^{-2i\phi_{B_s}} 
\end{equation}
How will it then be possible to test the predictions for the CKM
parameters shown in figs.~1? Let us look at each of them, in turn.

Since we do not expect eq.~(\ref{ratio}) to be valid, the mixing in
$B_d$ and $B_s$ might not be the right source of information on
$|V_{td}/V_{ts}|$. On the contrary, a clean way to measure $|V_{td}|$
is provided by the branching ratio of $K^+\rightarrow\pi^+\nu
\bar{\nu}$, which is not affected by gluino corrections. Our
prediction is shown in Table 2 for the four combined fits. Also not
affected by gluino corrections are the decay modes $B\rightarrow
\chi_s (\chi_d) \nu \bar{\nu}$ and, to a good approximation,
$B\rightarrow \chi_s (\chi_d) l^+ l^-$ \footnote{ $B\rightarrow \chi_s
(\chi_d)$ decays, as $\epsilon_K$, $B-\bar{B}$ mixing and $K\rightarrow
\pi$ decays, receive also corrections due to charged Higgs and
chargino-stop exchanges, controlled by the top Yukawa coupling, which
are there in any supersymmetric extension of the Standard Model. These
corrections are most prominent in $B\rightarrow \chi_s (\chi_d) l^+
l^-$, as in $B\rightarrow \chi_s (\chi_d) \gamma$, even for relatively
small $v_2/v_1$~\cite{gtcorr}. Not seeing any effect in $B\rightarrow
\chi_s l^+ l^-$ at $B$-factories would make this kind of correction
negligible everywhere else. In any case, they do not affect
${M_{12}(B_d) \over M_{12}(B_s)}$ nor the CP-asymmetries.} Therefore
the ratios $(B\rightarrow\chi_d)/ (B\rightarrow\chi_s)$ are another
potential source of information on $|V_{td}/V_{ts}|$, if
SU(3)-breaking effects, especially in $B\rightarrow \chi_s (\chi_d)
l^+ l^-$, can be kept under control.
\begin{table}
\centering
\renewcommand{\arraystretch}{1.1}
\begin{tabular}{||c||c|c|c|c||}
\hline
	& i)
	& ii)
	& iii)
	& iv) \\
\hline
\hline
	  BR$(K^+\rightarrow \pi^+ \nu\bar{\nu})/10^{-10}$
	& $0.98^{+0.40}_{-0.30}$
	& $0.83 \pm 0.20$
	& $0.97^{+0.50}_{-0.30}$
	& $0.84^{+0.17}_{-0.22}$ \\
\hline
	  BR$(K_L\rightarrow \pi^0 \nu\bar{\nu})/10^{-10}$
	& $0.22 \pm 0.13$
	& $0.17 \pm 0.07$
	& $0.20^{+0.08}_{-0.10}$
	& $0.14 \pm 0.05$ \\
\hline
\end{tabular}
\mycaption{Predictions for $K \rightarrow \pi \nu\bar{\nu}$ decays in the four ``combined fits'' described in the text.}
\label{tab:decays}
\end{table}

The new phases $\phi_{B_d}$ and $\phi_{B_s}$ defined in (\ref{phases})
modify the CP asymmetries of neutral $B$-decays into CP eigenstates. For
example, concentrating on the coefficient of the $\sin \Delta m t$
term in the time dependent CP asymmetries, one has~\cite{bigi}
\begin{equation}
\label{CPpsi}
a_{CP}(B_d \rightarrow \psi K_S) = -\sin 2 (\beta + \phi_{B_d}), \qquad
a_{CP}(B_s \rightarrow \psi \phi) = -\sin 2  \phi_{B_s}
\end{equation}
or, neglecting possible penguin contributions or assuming their
elimination by a proper isospin analysis,
\begin{equation}
 a_{CP}(B_d \rightarrow \pi \pi) = \sin 2 (\alpha - \phi_{B_d}).
\label{CPpi}
\end{equation}
Therefore, $\sin 2\alpha$ and $\sin 2\beta$ are not directly
accessible from the CP asymmetries in $B$-decays, nor is $\sin 2\beta$
obtainable from $\epsilon_K$, although required to be
non-vanishing due to (\ref{eps}). Yet a correlation between
$a_{CP}(B_d \rightarrow \psi K_S)$ and $a_{CP}(B_d \rightarrow \pi
\pi)$, as given in (\ref{CPpsi}) and (\ref{CPpi}), can still be
predicted at least if use is made of (\ref{u}). This
correlation is shown at 90\% C.L. in fig.~2 for arbitrary values of
$\phi_{B_d}$.

\begin{figure}
\centering
\setlength{\unitlength}{\textwidth}
\begin{picture}(0.55,0.4)
\put(0,0){\epsfig{file=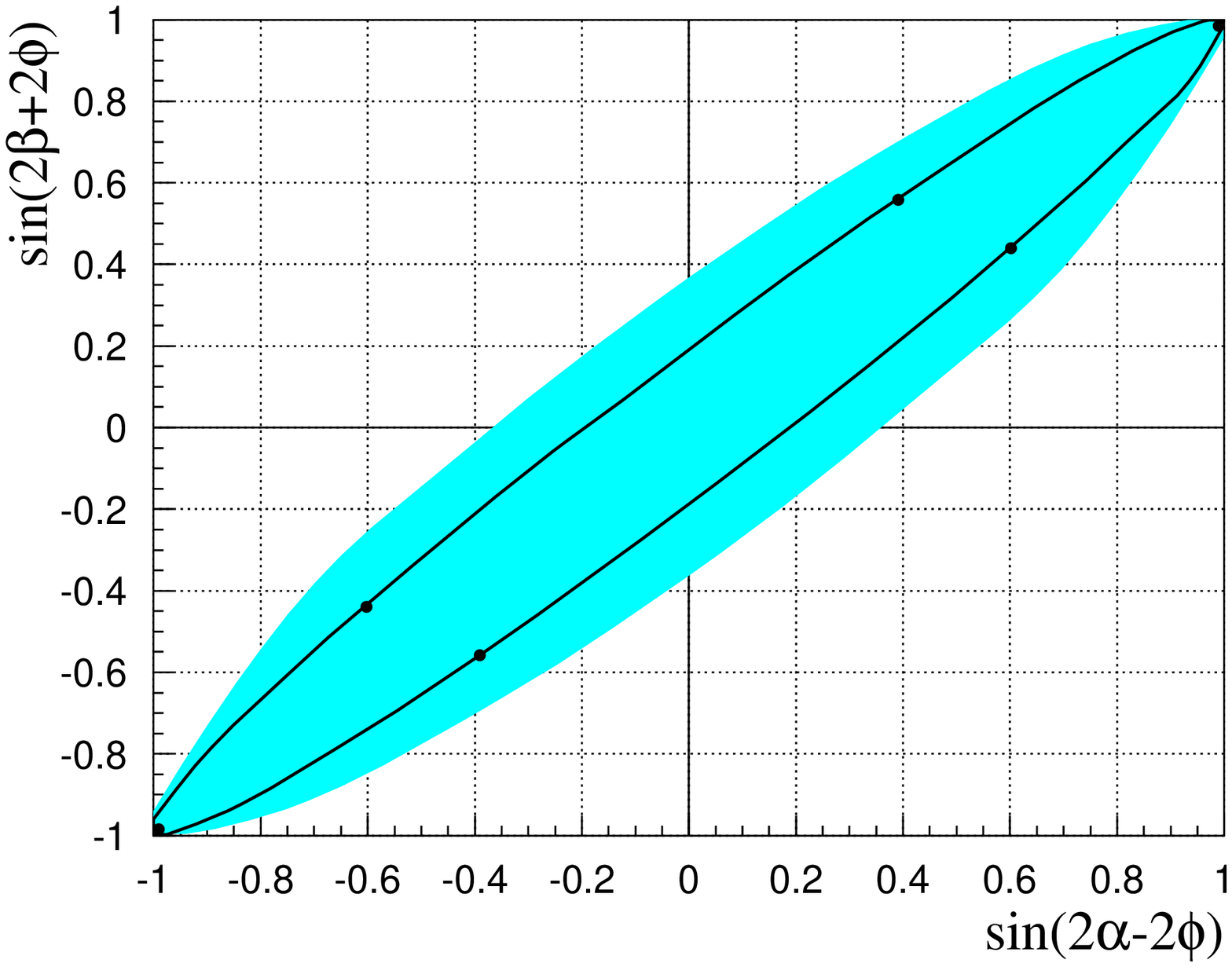,width=0.55\textwidth}}
\put(0.35,0.35){$A$}
\put(0.475,0.32){$B$}
\put(0.41,0.245){$C$}
\put(0.195,0.085){$D$}
\put(0.074,0.118){$E$}
\put(0.135,0.19){$F$}
\end{picture}
\mycaption{$90\%$ C.L. contour for the asymmetries measurable in
$B_d\rightarrow\psi K_S$ {\rm ($-\sin 2(\beta+\phi_{B_d})$)} and
$B_d\rightarrow \pi\pi$ {\rm ($\sin 2(\alpha-\phi_{B_d})$)} in fits
ii) and iv), which give indistinguishable results. A,D: $\phi=0,\pi/2$; B,E:
$\phi=\pi/6,2\pi/3$; C,F: $\phi=\pi/3,5\pi/6$.}
\label{fig:sinsx}
\end{figure}


A clean way to get $\sin^2\beta$ is again through $K$-decays. The
branching ratio for $K_L\rightarrow\pi^0\nu \bar{\nu}$ is short
distance dominated, it is not affected by gluino
corrections and is proportional, through a rather precisely known
coefficient~\cite{buras}, to
$|V_{td}V_{ts}|^2
\sin^2
\beta$. 
$K_L \rightarrow \pi^0\nu\bar{\nu}$ is predicted in Table 2. As shown
in fig.~1c, $\sin^2\beta$ is only weakly correlated with  
$\sin^2 \gamma$, measurable from the CP asymmetry in charged or
neutral $B$-decays into a final state, containing a $D$-meson and a
Kaon, which is not a CP-eigenstate~\cite{gam}. 

We expect that none of the CP asymmetries that we have mentioned so
far will be affected in any significant way by supersymmetric loop
corrections in the decay amplitudes, with the possible exception of
$a_{CP}(B_d \rightarrow \pi \pi)$. 
In this case, the dominant corrections should
come from supersymmetric penguin diagrams and would be erased from the
asymmetry by the isospin analysis invented to get rid of the (badly
known) gluon-mediated penguin diagrams of the SM. On the other hand,
asymmetries that could be significantly affected, in a controllable
way, by supersymmetric loop corrections in the decay amplitudes, from
penguin or box diagrams, are $a_{CP}(B_d \rightarrow \phi K_S)$ and
$a_{CP}(B_s \rightarrow \phi \eta)$. A new phase $\chi$ would be
introduced through the decay amplitude~\cite{newphase}:
\begin{equation}
\label{CPsss}
a_{CP}(B_d \rightarrow \phi K_S) = -\sin 2 (\beta + \phi_{B_d}
                +\chi), \qquad 
                a_{CP}(B_s \rightarrow \phi \eta) = -\sin 2 (
                \phi_{B_s}+\chi)
\end{equation}
that would make them deviate from $a_{CP}(B_d \rightarrow \psi K_S)$
and $a_{CP}(B_s \rightarrow \psi \phi)$ respectively, as given in
(\ref{CPpsi}).




\end{document}